# Structural attack to a pseudo-random generator

**Behrooz Khadem[1], Ali Madadi**

bkhadem@ihu.ac.ir , almadadi@ihu.ac.ir

**Imam Hussein Comprehensive University, Iran**

## Abstract

**Time, data and memory trade off attack is one of the most important threats against pseudo-random generators and resisting against it, is considered as a main criteria of designing such generators. In this research, the pseudo-random GMGK generator will be addressed and analyzed in details. Having indicated various weaknesses of this generator, we performed three different versions of structural attack on this generator and showed that proposed TMDTO attacks to this generator can discover blocks of plaintext with lower complexity than exhaustive search of space of key generator. Results indicated that the mentioned generator is lack of the security claimed by authors.**

## 1. Introduction

Today, satellite and cellular communication networks and internet play a main role in providing global universal computation and communication. Based on this, using communication tools in most cases resulted in sensitive and important data to be saved and accessed and transferred which in turn makes privacy security as a main issue in those networks. While several researches has been performed in fields like cryptography, security protocols and standards to access to secure communication tools and proportionate to application, there are problems in this regard yet to removing of which appropriate solutions must be devised.

Sequence codes is one of symmetrical cryptography algorithms that is highly important because of unique capabilities in communication and telecommunication networks like internet and GSM. Stream ciphers are used to make secure and verify electronic data transmission and are generally include a pseudo- random generator that generates a sequence of pseudo-random numbers or bits. For example those numbers, have been used in fuzzy cellular machines based on random numbers generators in [MSRB15].

Recently, several researches have been performed in field of pseudo-random generators in particular chaotic pseudo-random generators and various designs have been proposed. For further studies on design and security of chaotic generators see [AL06]، [PP10]، [FGBE13] and [K16].

Structural attacks are those attacks that don't concern the nature of cipher and its internal components and finds its weaknesses just through unique links among its inputs and outputs. Consequently, new information is detected on secure key or plain text. Time,

[1] Corresponding Author: Behrooz Khadem, Faculty of Information and Communication Technology, IHC university, Iran

memory tradeoff attacks TMTO are common structural attacks on symmetric ciphers. Those attacks are implementable in known (chosen)-plaintext attack model and known (chosen)-cipher text attack model, if available under certain conditions. One of the methods to reduce time of exhaustive key search attack in stream ciphers is TNDTO attack that enables the attacker to tradeoff optionally time, memory and data resources required for this attack together based on an equation named TMTO curve and as a result simplifies the attack. After Hellman proposed general principles for block ciphers [He80], several studies have also been performed on stream ciphers to adjust or generalize it. Those researches are best exemplified by works including results obtained by Babbage on stream ciphers [B95], Golic on A5 stream ciphers [G97], two considerable papers by Biryukov [BS00] and Biryukov et.al. introducing TMDTO attack and attack to A5/1 [BSW00], Dawson et.al., evaluating LILI-128 [DGMS00], Saarinen's work to attack LILI-128 [S02], Hong et.al., introducing new applications of TMDTO [HS05], Babbage et.al., for evaluating MICKEY [BD06], and Dunkelman et.al. [DK08], for introducing role of IV in TMDTO attack. Readers are recommended to study Krhovják et.al. In [KSN11] for quick review on theory and application of TMDTO attacks to stream ciphers.

This paper focuses mainly on the structural attack to a pseudo-random generator and is presented as follows. In Section 2 a new generator will be briefly described proposed in [GMGK16] after providing a general definition of pseudo-random generator. Also some of its most important weaknesses will be pointed out. Next three sections deal introduction of the three different structural attacks, particularly TMDTO briefly and then several attacks to the generator will be explained. The paper will be ended proposing summary of result and conclusion.

## 2. Summary of pseudo- random generator GMGK

Generally, a pseudo-random generator (statefull and synchronous) is generally shown by a quadruple $(X, S, \gamma, n)$ where X is space of input and output alphabet, S is the internal state space, $\gamma: S \rightarrow S \times X$ is the generator function (state transition) and n is the maximum number of output blocks for the generator (fig. 1) and is defined as[K16]:

$$\begin{cases} input: s_0 = Init(Key, IV) & . \\ (s_t, z_t) = \gamma_z(s_{t-1}), \quad (t = 1, \cdots, n) \\ output: \langle z_t \rangle & . \end{cases} \qquad (1.1)$$

In this generator during initializing step, the initial state $s_0$ is calculated with a function $Init$ selecting secure key $Key$ and a known initial value. Then, transition function for generator state $(\gamma_z)$ according to definition is iterated for producing $n_b$ blocks of key stream sequence $\langle z_t \rangle$. In a stream cipher, a block of cipher text is generated by combination of a block of plaintext and a block of key stream. In most cases of structural attacks to stream ciphers, the state transition function $(\gamma_z)$ or composite function $(\gamma_z O Init)$ is considered as one-way function that must be inversed.

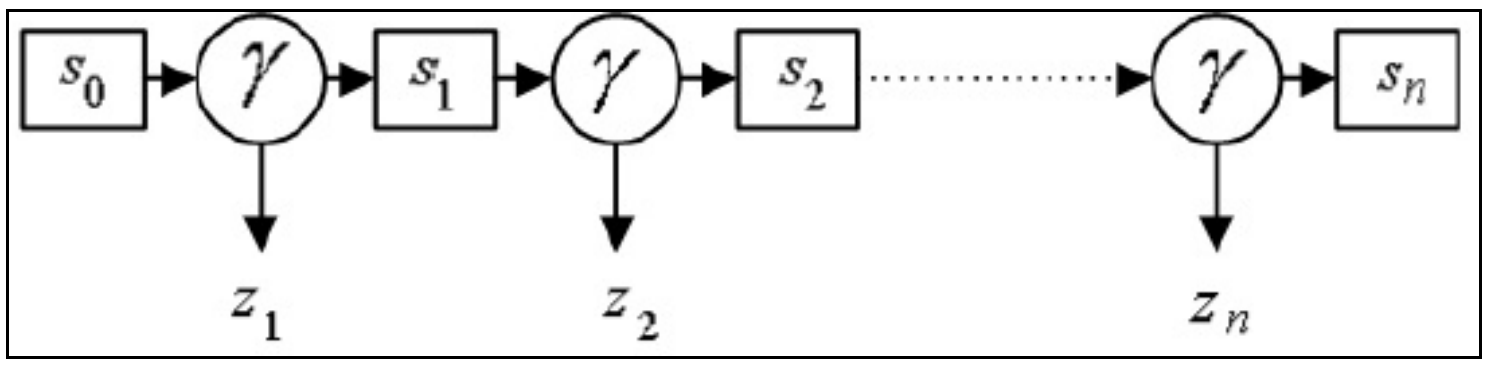

Figure 1: overview of a pseudo-random generator [K16]

### 2.1. A brief overview of GMGK and its main components

Gaeini et.al. have recently proposed a chaotic pseudo-random generator [GMGK16] (in this paper it is referred to as GMGK). This generator uses a 128 bits secure key $(Key)$, an initial value of 128 bits $(IV)$, and a 128 bits internal state $(L)$ and includes a linear shift register, 3 linear congruence generators (5,1) and 3 chaotic generator called Tent(1.2), and Logistic (1.3) and Chebyshev (1.4.). In congruence generators $(t > 0)$ and in chaotic maps $(t \geq 0)$, it was assumed that:

$$x_{t+1} = \begin{cases} \dfrac{x_t}{\alpha} & , \quad 0 \leq x_{t+1} < \alpha \\ \dfrac{1-x_t}{1-\alpha} & , \quad \alpha \leq x_{t+1} < 1 \end{cases} \qquad (1.2)$$

$$b_{tent}^t = \begin{cases} 0 \ , \ 0 \leq x_{t+1} < \alpha \\ 1 \ , \ \alpha \leq x_{t+1} < 1 \end{cases}$$

$$x_{t+1} = 3.9999\, x_t (1 + x_t)$$
$$b_{lo}^t = \begin{cases} 0 \ , \ 0 \leq x_{t+1} < 0.5 \\ 1 \ , \ 0.5 \leq x_{t+1} < 1 \end{cases} \qquad (1.3)$$

$$x_{t+1} = \mathrm{Cos}(4 * arc\,\mathrm{Cos}(x_t))$$
$$b_{cheb}^t = \begin{cases} 0 \ , \ \text{-}1 \leq x_{t+1} < 0 \\ 1 \ , \ 0 \leq x_{t+1} \leq 1 \end{cases} \qquad (1.4)$$

$$\begin{aligned} Z_1^t &= (b_{tent}^{t-1} + 1)(Z_3^{t-1} + Z_1^{t-1}) \bmod (2^{31} - b_{tent}^t) \\ Z_2^t &= (b_{lo}^{t-1} + 1)(Z_1^{t-1} + Z_2^{t-1}) \bmod (2^{31} - b_{lo}^t) \\ Z_3^t &= (b_{cheb}^{t-1} + 1)(Z_2^{t-1} + Z_3^{t-1}) \bmod (2^{31} - b_{cheb}^t) \end{aligned} \qquad (1.5)$$

In case of GMGK, pseudo- random numbers are generated in two steps. Those numbers are generated in the first step will be used as input of second step (fig.2).

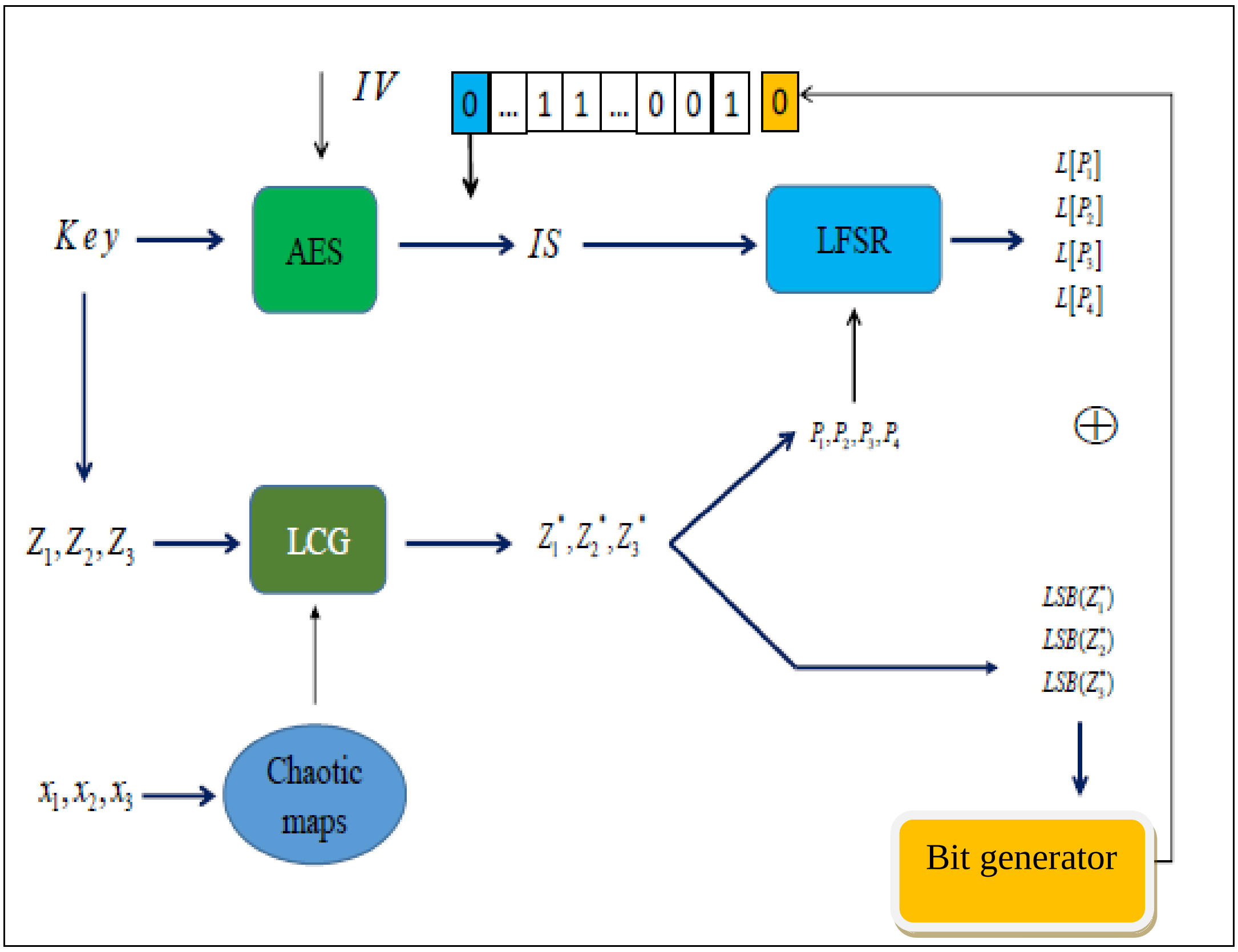

Fig 2: overview of GMGK generator [G16]

First step of algorithm is based on linear congruence generators and chaotic maps while the second step of algorithm is based on shift register whose feedback is formed in each step using algorithm of first step. In following sections, we will call the first step algorithm as setup algorithm and the second step algorithm as the generator algorithm. The structure of setup algorithm is as follow. Its input contains three values:

1. One 128 bits Initial value $(IV)$
2. One 128 bits key $(Key)$
3. Three initial values of chaotic maps used in (1.5)

In setup algorithm, the value of 128 bits $(IV)$ is encrypted as the plaintext block with $(Key)$ by AES and the block of output cipher text is considered as the initial state $(L)$ of shift register. From this time on, AES is not used elsewhere. Three initial values of chaotic maps according to (1.2), (1.3) and (1.4) are entered in chaotic generators and generate three pseudo-random (chaotic) bits of $b^0_{tent}$ ,$b^0_{lo}$ and $b^0_{cheb}$.

To calculate (1.6), the initial value $Z^0_1$ has been determined as first 32 bits, $Z^0_2$ as the second 32 bit and $Z^0_3$ as the third 32 bits of $(Key)$ and numbers $Z^t_1, Z^t_2, Z^t_3$ are generated according to (1.5). Then, values of $P^t_1, P^t_2, P^t_3, P^t_4$ are obtained based on (1.6). The symbol $(<< n)$ introduces n shift bits to right. In the following section, an output bit is generated according to (1.7) that makes the output sequence (key stream) of the generator algorithm.

$$
\begin{aligned}
P_1^t &= (Z_1^t << (1+b_{lo}^t)) \mod 128 \\
P_2^t &= (Z_2^t << (1+b_{cheb}^t)) \mod 128 \\
P_3^t &= (Z_3^t << (1+b_{tent}^t)) \mod 128 \\
P_4^t &= \min\{Z_1^t, Z_2^t, Z_3^t\} << \\
&\quad (b_{tent}^t + b_{lo}^t + b_{cheb}^t) \mod 128
\end{aligned}
\tag{1.6}
$$

$$
\begin{aligned}
z_t = LSB(Z_1^*) \oplus LSB(Z_2^*) \oplus LSB(Z_3^*) \oplus \\
L[P_1] \oplus L[P_2] \oplus L[P_3] \oplus L[P_4]
\end{aligned}
\tag{1.7}
$$

In (1.7), $LSB(Z_i^t)$ indicates the lowest value bit in $Z_i^t$ and $L[P_i^t]$ indicates the bit value that is internal state of $L$ in $P_i^t$-th position. The output value of the generator algorithm, ($z_t$) is the final generated bit that is added to the right side of internal state $L$ when the generator algorithm is executed and consequently the last 128 bits are considered as a new internal state. This process continues to have all bits of key stream to be generated. After all stream key sequence is generated, the first L bits of it (that has been generated once by AES) will be disposed. Together with initial values of chaotic map, (*Key*) is transmitted to the transmitter through a private channel and ($IV$) is transitioned through the public channel at first of key stream to enable the receiver to decrypt data.

### 2.2. The main weaknesses of GMGK

Weaknesses of this generator are classified into two classes; weaknesses of chaotic generator and structural weaknesses:

- Failing to use standard discretization method in chaotic map that causes the resulted sequence to lose its chaotic properties (random) after a short term [K16].
- Uncertain bit size of initial values of chaotic map used in setup algorithm for the receiver that makes calculation of key length difficult.
- The computational accuracy of machine when calculating values of chaotic sequences isn't expressed explicitly that makes decryption difficult.
- Size of $IV$ is smaller than 3/2 of $Key$ size that indicates the generator isn't able to resist against TMTO attack [DK08].
- Size of ($L$) isn't bigger than ($Key$) that indicates the generator acts weak against TMTO attack [G97].

## 3. Types of structural attacks to current ciphers

The basic idea of these attacks is belong to Hellman and has proposed two decades ago for block ciphers [He80] that generalized by Biryukov and Shamir proposed to stream ciphers in [BS00] and they called it TMDTO. The basic idea in this attack is that the adversary attempts to find a one-way function and a heuristic method to inverse it effectively and to reduce time complexity for discovering secure key. This attack is always performed in two steps as offline (pre-processing) and online (main). In case of stream ciphers, the pre-processing step first calculates the output corresponding to a number of inputs of one-way function and then these pairs of inputs and outputs are saved in a relatively large table. The time and memory of preprocessing step is not important and is generally neglected. Then in the main step, the adversary searches the output which is obtained from a small piece of real key stream sequence to find at least one of the saved outputs. As a correspondence is detected between real output of the generator (key stream sequence) and one of predetermined outputs, the input related to this output is obtained from table and the adversary is allowed to generate complete output key stream sequence of the generator and to discover remaining blocks of plaintext using sequence of cipher text blocks. Before studying proposed attacks, we introduced some notations which are used in this paper in table 1.

Table 1: notations

| **Symbol** | **Concept** |
|---|---|
| $f$ | Concerned one-way function |
| N | Size of search space |
| P | Time of execution of pre-processing step |
| M | Memory of execution of pre-processing step |
| D | Number of $f$ outputs owned by the adversary |
| T | Time needed for executing the main step |

Babbage, Golic, Biryukov and Shamir have used separately TMDTO attack for a pseudo-random generator in a stream cipher. In this attack, they all have considered a function that maps the internal state space of the generator on a piece of key stream sequence. Having inversed this function, they discovered main part of the plaintext. The important notes of TMTO curve and related conditions in these attacks are presented in table 2.

## 4. First proposed attack

In case of GMGK, the size of internal state is equal to the size of secure key. In various researches this weakness have considered as one of insecurity factors against a TMDTO attack. We used this weakness to implement our first proposed attack. Now we describe the attack scenario.

Let $S = \{0,1\}^{128}$ is the set of possible states of GMGK and $f : S \to Y$ is a function that maps each state of $S$ to the first 128 bits of a key stream sequence produced from that state.In the preprocessing step, the adversary wishes to generate a large table in two columns (of 128 bits) and $m$ rows (table 3). In order to create each row, it selects a 128 bits randomly (or intelligently in some of advanced methods) as an internal state value of GMGK and places it in first column ($s_i$) and then 128 times executes function $f$ iteratively on it to generate an output of 128 bits (key stream sequence). It places the output in the second column $y_i = \langle b_0 \cdots b_{127} \rangle_i$ and saves in the table in the ascending order of outputs. In the main step, (supposing that secure key is fixed) the adversary has $D$ pieces of a real key stream sequence and searches for similar piece for each of them in second column of table 3. Provided that $D$ and $m$ are large enough, regarding to the birthday problem, the attack success probability will be large and increasing number of columns of this table will increase this probability. In case that it succeeds to find such piece it will determine the internal state corresponding to it as the targeted internal state and using it and repeating iteratively $f$ can generate rest of bits of key stream. Consequently, it will discover a large part of plain text blocks.

Table 2: The important notes of previous attacks

| **Reference** | **TMTO curve and related conditions** | **Size of a sample value suitable on TMTO curve** |
|---|---|---|
| [B95] | $TM = N$,<br>$D = N/M$,<br>$P = \sqrt{N}$ | $T = M = D = \sqrt{N}$ |
| [G97] | $TM = N$,<br>$1 \le T \le D$,<br>$P = O(M)$ | $T = M = D = \sqrt{N}$ |
| [BS00] | $TM^2D^2 = N^2$,<br>$1 \le D^2 \le T \le N$,<br>$P = N/D$ | $P = N^{2/3}, T = N^{2/3}$,<br>$M = N^{1/3}, D = N^{1/3}$ |
| [BSW00] | $TM^2D^2 = N^2$,<br>$D^2 \le N$,<br>$P = N/D$ | $T = M = \sqrt{N}$,<br>$D = N^{1/4}$ |
| [BSW00] | $TM^2(DR)^2 = (NR)^2$<br>$(DR)^2 \le T \le N$ ,<br>$P = N/D$ | |

Table 3: generated memory during preprocessing step

| $i$ | $s_i$ | $y_i = \langle b_0 \cdots b_{127} \rangle_i$ |
|---|---|---|
| 1 | $s_1$ | $y_1$ |
| 2 | $s_2$ | $y_2$ |
| …… | …… | …… |
| $m$ | $s_m$ | $y_m$ |

In the main step, (supposing that secure key is fixed) the adversary has $D$ pieces of a real key stream sequence and searches for similar piece for each of them in second column of table 3. Provided that $D$ and $m$ are large enough, regarding to the birthday problem, the attack success probability will be large and increasing number of columns of this table will increase this probability. In case that it succeeds to find such piece it will determine the internal state corresponding to it as the targeted internal state and using it and repeating iteratively $f$ can generate rest of bits of key stream. Consequently, it will discover a large part of plain text blocks.
Based on ideas discussed by Biryukov and Shamir in [BS00] , this attack is possible with TMTO curve $TM^2D^2 = N^2$ . Regarding the value of $N = 2^{128}$ and 3rd row of table 2, complexity of the abovementioned attack can be seen from 1st column of table 6. As seen, all resources of attack are separately smaller than search space of secure key. It indicates that the mentioned generator lacks 128 bits of security claimed by authors.

## 5. Second proposed attack

While Values of complexity obtained in the first step are smaller than that of exhaustive search of secure key space, it is probably impossible functionally. Because of this, we describe a second attack to GMGK generator. Biryukov, Shamir and Wagner indicate a weakness related to Rivest sampling method in TMDTO attack and proposed a new sampling method (BSW method in this paper) [BSW00]. In this method, each of outputs of the one-way function with a certain state (e.g. starts with k zero bits) is called special output and the internal state corresponding to it, is called special state. Its idea was to restricting the memory of preprocessing step to special states and outputs so that in the main step it was sufficient that only these outputs to be examined. Therefore, memory and time required for preprocessing and main attack time used to decrease with a known factor. They found that because the internal state in most of stream ciphers changes slightly after each output bit generated, this type of attack succeeds more likely, in practice. Babbage and Dad in [BD08] used this type of attack.
Let $S = \{0,1\}^{128}$ be the set of all possible states of GMGK and $f : S' \to Y'$ where $S' \subset S$ and $Y' \subset Y$, and $S'$ is the set of those 128 bits which their least significant 64-bits are different.
In the preprocessing step, the adversary also wishes to generate a large table consisting of two 128 bits columns and $m$ rows (table 4). But instead of random selection of 128 bits as the internal state of GMGK $s_i'$ , the adversary here just generates those 128 bits which their least significant 64-bits are different and then executes function $f$ 128 times iteratively on each of them to generate a 128 bit output $y_i' = \langle b_0 \cdots b_{127} \rangle_i$ and saves in the table in the ascending order of outputs.
Since each internal state of GMGK generator, shifts one bit left side in each cycle and the generated bit is added to its right most place, it is seen that each two sequential states have a set of common 127 bits and also each 128 sequential states have at least a set of 64 common bits, so it will be sufficient to save an index member of each set to reduce the search memory and time. Thus, our second attack will decrease time and memory of preprocessing step of attack with a reduction factor of $R = 2^{-8}$.
Suppose the adversary have $D$ pieces of a real key stream sequence in this step, so in the main step, she seeks for each of them a target state in table 4. Again provided that $D$ and $m'$ are large enough, regarding to the birthday problem, the attack success probability will be large and increasing number of columns of this table will increase this probability... If the adversary could find such a piece, she will be able to find a target internal state form table 4 and using it she can generate remaining of the real key stream sequence by repeating iteratively $f$ and finally discovers the main part of plaintext as a result.

Table 4: generated memory during preprocessing step

| $i$ | $s'_i$ | $y'_i = \langle b_0 \cdots b_{127} \rangle_i$ |
|---|---|---|
| 1 | $s'_1$ | $y'_1$ |
| 2 | $s'_2$ | $y'_2$ |
| …… | …… | …… |
| $m'$ | $s'_{m'}$ | $y'_{m'}$ |

The second proposed attack is practical based on BSW sampling method. Regarding one of sample points of this curve with values contained in 5th row of table 2, the complexity of the abovementioned attack can be observed in 2nd column of table 6. As seen from table 6, all used resources of the attack are separately smaller than the size of search space for secure key of GMGK and they indicate that the mentioned generator doesn't have 128 security bits that claimed by authors.

## 6. Third proposed attack

This section deals third proposed attack to GMGK and will be discussed in two steps. Let $f : S \to S$ be one-way function and the adversary wishes to make its inverse from $s_{i+1}$ to $s_i$. In the preprocessing step, if the adversary has $\log_2 N$ bits of a real key stream sequence, then she will be able to search for one of intermediate states of the generator in the state for known plain text [BSW00]. Supposing that $b_1 \cdots\cdots b_j$ is a piece of real key stream sequence, the adversary creates a table in preprocessing step and occupies $m$ input state $s_i$ and calculates with a known value $IV$ and a large number of random keys $Key$, their corresponding $s_{i+1}$ output states saves in the table in the ascending order of outputs.

Table 5: created memory in preprocessing step

| $i$ | $s'_i$ | $u_i = f(s_i)$ |
|---|---|---|
| 1 | $s_1$ | $u_1$ |
| 2 | $s_2$ | $u_2$ |
| …… | …… | …… |
| $D$ | $s_D$ | $u_D$ |
| …… | …… | …… |
| $M-D$ | $s_{M-D}$ | $u_{M-D}$ |
| …… | …… | …… |
| $M$ | $s_M$ | $u_M$ |

In the main step, the adversary has a real key stream sequence of $D(\log_2 N+1)$ bits. She considers each piece of $b_1 \cdots\cdots b_j$ from this sequence (called a window). Then she considers $u_1 \cdots\cdots u_j$ in table 5 and compares theirs corresponding output bits with $b_1 \cdots\cdots b_j$.
Since conditions of the proposed attack conforms to conditions of the 3rd column of table 2, the adversary will be able to find at least one window coinciding to sequence $b_1 \cdots\cdots b_j$. Having find this window and referring to table 5, the adversary achieves the state placed in 1st row of this window which is her target state.

The third proposed attack is practical with TMTO curve $TM^2D^2 = N^2$. The complexity of the attack is seen from 3rd column of table 6 regarding 5th row in table 2. The attack time complexity is considerably lower than that of exhaustive search of key space and thus is considered as a successful attack. It can be indicated that if size of input data doubles (key stream sequence ) then time complexity of the preprocessing step will decrease to half and complexity of processing step to quarter.

## 7. Summary and conclusions

This paper evaluated and analyzed the GMGK pseudo-random generator in details. We indicated weaknesses of this generator as well as three versions of different structural attacks to it and indicated that proposed TMDTO attacks to it can discover a large part of the plain text with lower complexity compared to exhaustive search of key space which in turn demonstrates that the mentioned generator lacks the security claimed by the authors. Table 6 summarize results of data, memory, and time complexity for three above proposed attacks. It allows the reader to use one of those methods to attack this generator proportionate to his or her resources and computational tools.

Table 6: complexity of proposed attacks

| Parameter | 1st attack | 2nd attack | 3rd attack |
|---|---|---|---|
| $N$ | $2^{128}$ | $2^{128}$ | $2^{128}$ |
| $P$ | $2^{42.6}$ | $2^{48}$ | $2^{48}$ |
| $M$ | $2^{85.2}$ | $2^{52}$ | $2^{64}$ |
| $D$ | $2^{85.2}$ | $2^{40}$ | $2^{32}$ |
| $T$ | $2^{42.6}$ | $2^{40}$ | $2^{64}$ |